\documentstyle[12pt]{article}
\def \beq{\begin{equation}}
\def \cn{Collaboration}

\def \eeq{\end{equation}}
\def \es{E$_6$}
\def \evt{$e^+ e^- \gamma \gamma E\!\!\!/_T$}

\def \sso{\sigma/\sigma_0}

\def \sub{SU(2)$_I \times$ SU(6)}
\def \sui{SU(2)$_I$}

\begin{document}
\renewcommand{\thetable}{\Roman{table}}
\begin{titlepage}
\vspace{-3in}
\rightline{CERN-TH/96-209}
\rightline{EFI-96-30}
\rightline{hep-ph/9607467}
\begin{center}
{\large\bf An E$_6$ interpretation of an $e^+ e^- \gamma \gamma E\!\!\!/_T$
event\footnote{Submitted to Physical Review D.}} \\
\vspace{1.5cm}
{\large Jonathan L. Rosner} \\
\vspace{.5cm}
{\sl CERN, 1211-CH Geneva 23, Switzerland} \\
\vspace{.5cm}
{\sl Enrico Fermi Institute and Department of Physics}\\
{\sl University of Chicago, Chicago, IL 60637 USA}
\footnote{Permanent address.}\\
\vspace{1.5cm}
\begin{abstract}
The lowest-dimensional representation of the group E$_6$ contains both the
standard quarks and leptons and a set of exotic quarks and leptons whose decays
can involve a series of chains ending in radiative decay of one light neutrino
species to another.  An example is given based on the decomposition E$_6 \to$
SU(2)$_I \times$ SU(6), where SU(2)$_I$ is an ``inert'' subgroup whose gauge
bosons $W_I^{(\pm)}$ and $Z_I$ are all electromagnetically neutral, while SU(6)
contains the conventional SU(5) grand-unified group. The possibility is
explored that such a chain is responsible for an event observed by the Collider
Detector at Fermilab (CDF) involving the production in proton-antiproton
collisions at $E_{\rm c.m.} = 1.8$ TeV of an electron-positron pair, two
photons, and missing energy ($e^+ e^- \gamma \gamma E \!\!\!/_T$). 
\end{abstract}

\end{center}
\leftline{PACS codes: 14.70.Pw, 12.60.Cn, 12.10.Dm, 13.38.Dg}
\vfill
\leftline{CERN-TH/96-209}
\leftline{July 1996}
\end{titlepage}

\centerline{\bf I.  INTRODUCTION}
\bigskip

The ``grand unification'' of strong and electroweak interactions in a larger
symmetry, and the identification of quarks and leptons as objects related to
one another under this symmetry, involves such groups as SU(5) \cite{GG},
SO(10) \cite{SO}, and \es~\cite{E6}.  We briefly recall some properties of
each group. 

Within SU(5) a specific choice of representations (${\bf 5}^* + {\bf 10}$) is
required for the left-handed fermions in order to accommodate the known states
and to eliminate anomalies. This choice is automatic if left-handed fermions
are assigned to the ${\bf 16}$-dimensional spinor multiplet of SO(10); the
additional state is a right-handed neutrino. Anomalies are not present in
SO(10), as long as matter belongs to complete multiplets.

The lowest-dimensional representation (${\bf 27}$) of the group \es~contains
the ${\bf 16}$ of SO(10), as well as ${\bf 10}$- and ${\bf 1}$-dimensional
(``exotic'') representations of SO(10).  There has been some interest in \es~as
a result of its appearance in certain versions of superstring theories
\cite{E6SS,rev}.

In the present article we discuss some properties of a decomposition
\cite{RR,LR} of \es~into a subgroup \sub, where the subscript $I$ stands for
``inert.''  The SU(6) contains the conventional grand-unified group SU(5) and
an additional U(1) factor which may be denoted U(1)$_{51}$.  The gauge bosons
of SU(2)$_I \times$ U(1)$_{51}$ are all electromagnetically neutral.  These
gauge bosons may mediate some interesting processes in hadronic collisions,
electron-positron annihilations, and $e^- p$ reactions. 

We have been stimulated to recall features of the present \es~decomposition by
the Collider Detector at Fermilab (CDF) Collaboration's report \cite{eegg} of
an event with an electron-positron pair, two photons, and missing energy
(\evt), produced in proton-antiproton collisions at $E_{\rm c.m.} = 1.8$ TeV. 
Alternative interpretations of this event have appeared within the context of
supersymmetry \cite {sseegg} and in one non-supersymmetric model \cite{BM}. 
There is still a need for extensive discussions of standard-model backgrounds
to this event, such as multiple interactions, radiative production of $W$
pairs, effects of cracks in the detector, and so on.

While we are aware of the dangers of speculations based on a single event, the
possibility that one is seeing evidence for an extended gauge structure (such
as occurs in \es) is sufficiently appealing and predictive that it is worth
considering at present, even though many of the predictions have been in the
literature for some time. Our picture will be explicitly non-supersymmetric and
is meant in part to illustrate the pitfalls of too hasty a conclusion that a
given class of events has proved the validity of low-energy supersymmetry. 

In Section II we recall some of the necessary \es~group theory.  Implications
for the CDF \evt~event and others produced in hadron colliders are treated in
Sec.~III.  Some signatures in other machines are noted in Sec.~IV, while Sec.~V
concludes.
\newpage

\centerline{\bf II.  \es~DECOMPOSITION}
\bigskip

\leftline{\bf A.  Multiplet structure}
\bigskip

The ${\bf 27}$ of \es~corresponding to the first family of left-handed quarks
and leptons may be decomposed in the following manner under \sub: 
\beq \label{eqn:decomp}
({\bf 2}_I,{\bf 6}^*)_L = \left[ \begin{array}{c c}
                    \bar h_1 & \bar d_1 \\
                    \bar h_2 & \bar d_2 \\
                    \bar h_3 & \bar d_3 \\
                     \nu_E   &  \nu_e   \\
                      E^-    &   e^-    \\
                    \bar N_e &   n_e    \\ \end{array} \right],~
({\bf 1}_I,{\bf 15})_L = \left[ \begin{array}{c c c c c c}
     0    &  \bar u_3 & -\bar u_2 & d_1 & u_1 &   h_1   \\
-\bar u_3 &     0     &  \bar u_1 & d_2 & u_2 &   h_2   \\
 \bar u_2 & -\bar u_1 &     0     & d_3 & u_3 &   h_3   \\
  -d_1    &   -d_2    &   -d_3    &  0  & e^+ & \bar N_E \\
  -u_1    &   -u_2    &   -u_3    & -e^+ & 0  &   E^+   \\
  -h_1    &   -h_2    &   -h_3    & -\bar N_E & -E^+ & 0 \\
\end{array} \right].
\eeq
Similar decompositions hold for the second and third quark-lepton families.

Although the exotic fermions in \es~have been discussed previously (see, e.g.,
\cite{E6} and \cite{zp}), we review them briefly.  We mention the properties of
the left-handed states; those of the right-handed states may be obtained via
CP-conjugation.
\begin{itemize}

\item $h$ is a weak-isosinglet quark with charge $-1/3$.

\item $\nu_E$ and $E^-$ are a weak isodoublet; so are $E^+$ and $\bar N_E$.  We
write $\bar N_E$ rather than $\bar \nu_E$ to stress the possibility that
$\nu_E$ and $\bar N_E$ may be two distinct Majorana neutrinos rather than
components of a single Dirac neutrino. 

\item $\bar N_e$ is the left-handed antiparticle (the CP-conjugate) of the
right-handed neutrino $N_e$.  As in the previous case, $\nu_e$ and $\bar N_e$
may be two distinct Majorana neutrinos rather than components of a single Dirac
neutrino. 

\item $n_e$ is a Majorana neutrino which is a singlet under both left-handed
and right-handed SU(2).

\end{itemize}

All the exotic fermions listed above except $n_e$ may be assigned to a {\bf
10}-plet of SO(10) under \es $\to$ SO(10) $\times$ U(1).  The $n_e$ may be
assigned to a singlet of SO(10).  An alternative assignment to SO(10)
multiplets is generated by interchanging states in the two columns of
$({\bf 2}_I,{\bf 6}^*)_L$ \cite{EW,EMa}.

With the above descriptions it should be clear how subgroups of SU(6) such as
color SU(3) and weak (left-handed) SU(2) act on the multiplets in
Eq.~(\ref{eqn:decomp}).  For example, in the multiplet $({\bf 2}_I,{\bf
6}^*)_L$, color SU(3) acts on the first three rows, while SU(2)$_L$ acts on the
fourth and fifth rows.  The conventional grand-unified SU(5) acts on the first
five rows.  The behavior of SU(6) subgroups acting on the {\bf 15} is best
seen by constructing it as the antisymmetric product of two {\bf 6}'s.  Thus,
$(u_i,d_i)_L~(i=1,2,3)$ and $(E^+,\bar N_E)$ form SU(2)$_L$ doublets.
\bigskip

\leftline{\bf B. U(1) charges in SU(6) $\to$ SU(5) $\times$ U(1)}
\bigskip

The simplest pattern of subsequent breakdown after \es~$\to$ \sub~is SU(6)
$\to$ SU(5) $\times$ U(1)$_{51}$, where SU(5) is the conventional grand-unified
group and U(1)$_{51}$ denotes an extra U(1) factor.  Adopting integral values
for the charges $Q_{51}$ of this U(1), we may decompose the {\bf 6}$^*$ of
SU(6) in Eq.~(\ref{eqn:decomp}) as ${\bf 6}^* = {\bf 5}^*_1 + {\bf 1}_{-5}$
and, since a {\bf 15} is the antisymmetric product of two {\bf 6}'s, we find
${\bf 15} = {\bf 10}_{-2} + {\bf 5}_4$.  Here the bold-face numbers on the
right denote the dimension of the SU(5) representation, while the subscripts
denote the U(1) charges $Q_{51}$. 
\bigskip

\leftline{\bf C.  Fermion masses}
\bigskip

We seek a pattern of mass splittings consistent with the hypothesis that all
the exotic fermions which can couple to the photon and $Z$ have masses large
enough that they will not have been produced in the tens of millions of $Z$
decays observed at the CERN LEP electron-positron collider and in the smaller
amount of data collected at higher energies. The mass splittings will be
implemented by means of Higgs bosons belonging to a {\bf 27}-plet of \es,
through the \es-invariant trilinear coupling of three {\bf 27}'s. 

The similarity of Higgs and fermion representations is a feature which makes
\es~particularly appealing in supersymmetric theories. Thus, without making any
necessary claims of supersymmetry, we will use a tilde to denote a scalar
particle transforming in the same manner under \es~or \sub~as the neutral
states in Eq.~(\ref{eqn:decomp}).  The Higgs bosons, their transformation
properties, and the effects of their vacuum expectation values (vevs) are
listed in Table I. 

\begin{table}
\caption{Higgs bosons belonging to the {\bf 27}-plet of \es~and their
transformation properties under some of its subgroups.}
\begin{center}
\begin{tabular}{c c c c l} \\ \hline
Boson          & $I_{3L}$ & $I_{3I}$ & $Q_{51}$ & What its vev does \\ \hline
$\tilde \nu_E$ &   1/2    &    1/2   & 1 & Gives $d,e$ Dirac mass \\
$\tilde \nu_e$ &   1/2    &  $-1/2$  & 1 & Mixes exotics, non-exotics \\
$\tilde{\bar N_e}$ & 0     &    1/2   & $-5$ & Mixes exotics, non-exotics \\
$\tilde n_e$   &    0     &  $-1/2$  & $-5$ & Gives $h,\nu_E,E$ Dirac mass \\
$\tilde {\bar N_E}$ & $-1/2$ &    0    & $4$ & Gives $u,\nu$ Dirac mass \\
\hline
\end{tabular}
\end{center}
\end{table}

The ``standard'' Higgs bosons in the present notation are $\tilde \nu_E$ and
$\tilde{\bar N_E}$.  Sufficiently large Dirac masses for the exotic fermions
$h$, $\nu_E$, and $E$ may be generated by a vev of the boson $\tilde n_e$. 
Such a Dirac mass term couples $\nu_E$ with $\bar N_E$.  Exotic fermions may be
mixed with non-exotic ones via vevs of the two remaining Higgs bosons $\tilde
\nu_e$ and $\tilde{\bar N_e}$.  These vevs may be very small if some selection
rule forbids the mixing of exotic and non-exotic fermions. Thus, a reasonable
hierarchy for vevs would be 
\beq \label{eqn:hie}
\langle \tilde n_e \rangle  = {\cal O}({\rm TeV}) \gg
(\langle \tilde \nu_E \rangle,~\langle \tilde{\bar N_E} \rangle)
= {\cal O}(v) \gg
(\langle \tilde \nu_e \rangle,~\langle \tilde{\bar N_e} \rangle)~~~,
\eeq
where $v = 246$ GeV $ = 2^{-1/4} G_F^{-1/2}$ characterizes the electroweak
breaking scale.

As mentioned in Ref.~\cite{JRE6}, one can describe all fermion masses
satisfactorily using the pattern suggested by Table I and employing
\es-invariant couplings, with the exception of neutrinos.  Since Dirac masses
for up-type quarks and neutrinos both arise through the vev of the Higgs boson
$\tilde{\bar N_E}$, one needs (i) to introduce some additional source of a
large Majorana mass for $\bar N_e$ (see, e.g., \cite{NS}), thereby causing
ordinary neutrinos to have very small Majorana masses \cite{seesaw}, (ii) to
provide an additional singlet of \es~with which $\bar N_e$ can form a Dirac
mass \cite{EW}, or (iii) to explicitly forbid the trilinear coupling between a
pair of fermions transforming as $({\bf 2}_I,{\bf 6}^*)_L$ and a boson
transforming as $({\bf 1}_I,{\bf 15})_L$.  We shall adopt the last point of
view, since a fairly light $\bar N_e$ will play a likely role in our
explanation of the \evt~event. We regard this as the least satisfactory feature
of the present model. 

There appears to be no phenomenological need to generate a mass for $n_e$,
and no source of such a mass except through the couplings $n_e \nu_E
\tilde{\bar N_E}$ or $n_e \bar N_E \tilde \nu_E$ (whose effects could be well
overwhelmed by a Dirac mass involving the pairing of $\bar N_E$ with $\nu_E$). 
Thus {\it an appealing candidate for a light state is the state
$n_e$,} as has been pointed out elsewhere \cite{EMa,Rizzo,EM,sterile}.

The Dirac masses of the exotic fermions $h$, $E$, and $\nu_E$ could be of any
values high enough to evade bounds associated with $Z$ decays and with more
recent higher-energy electron-positron collision experiments at LEP.
As in the case of $b$ and $\tau$, masses which start out identical at very
small distance scales will evolve at larger distances as a result of differing
gauge interactions in such a way that one will expect exotic quarks to be more
massive (perhaps by roughly the factor $m_b/m_\tau$) than exotic leptons.
\bigskip

\leftline{\bf D.  Exotic gauge boson masses and couplings}
\bigskip

We assume that in the breakdowns \es $\to$ \sub~and SU(5) $\to$ SU(3)$_c
\times$ SU(2)$_L \times$ U(1)$_Y$ (where $Y$ is the standard weak hypercharge)
the gauge bosons corresponding to the broken symmetries obtain super-heavy
masses.  Thus, we are left with the gauge bosons of SU(2)$_I \times$
U(1)$_{51}$ to discuss. 

In the hierarchy (\ref{eqn:hie}), the largest vev is acquired by a doublet of
\sui~with non-zero charge $Q_{51}$.  This situation is very close to that of
the Weinberg-Salam model.  If this were the only source of SU(2)$_I \times$
U(1)$_{51}$ breaking, we would have three massive bosons (two lighter than the
third) and a massless boson.  For simplicity, we assume instead that the
U(1)$_{51}$ factor is broken at a high mass scale by some other mechanism and
that we have only to deal with \sui. In that case we will have a theory
equivalent to the Weinberg-Salam model with $\theta = 0$, and there will
be three electromagnetically neutral bosons, each with mass of several hundred
GeV.  (A lower limit of order 10$^5$ GeV on the scale of \sui~breaking was
obtained in \cite{CM} with specific model-dependent assumptions and does not
apply here.)

We use the notation $W_I^{(\pm)}$ for two of the neutral bosons to denote the
fact that they change $I_{3I}$ by $\pm 1$ unit. The third boson (which couples
to $I_{3I}$ but does not change it) will be denoted by $Z_I$.  The masses of
the three bosons will be 
\beq
M_I = g_I V /2~~,~~~V^2 \equiv \sum_i \langle \tilde n_i \rangle^2~~~,
\eeq
where $g_I$ is the \sui~coupling constant (probably no stronger than the
standard SU(2)$_L$ electroweak coupling constant) and the sum is over all
families of Higgs bosons transforming as $\tilde n_e$.  $V$ is likely to be a
number of order 1 TeV if the exotic fermions discussed above are to be
responsible for signals observed in present collider experiments.  The
possibility of a second $Z'$ within \es, if one does not choose to break the
U(1)$_{51}$ symmetry at some high mass scale, should be kept in mind. 
\bigskip

\centerline{\bf III. EFFECTS OF $W_I$ AND $Z_I$ AT HADRON COLLIDERS}
\bigskip

Some features of exotic fermion production and decay via gauge interactions
mediated by $W_I$ and $Z_I$ were discussed in \cite{LR}.  We concentrate in
this section on production via $d \bar d$ collisions and decay via $W_I$
exchange.
\bigskip

\leftline{\bf A.  Production and decay of $Z_I$}
\bigskip

The states coupling to $Z_I$ are the members of the $({\bf 2}_I,{\bf 6}^*)_L$
in Eq.~(\ref{eqn:decomp}).  Each state couples with equal strength, since each
has $I_{3I} = \pm 1/2$.  The $Z_I$ can be produced in the direct channel in
electron-positron collisions, or it can be produced in hadronic collisions
via the $d \bar d \to Z_I$ subprocess.  Since $d$ quarks are softer than $u$
quarks in a proton (and there are fewer of them), the production of $Z_I$ at
the Fermilab Tevatron (involving proton-antiproton collisions) will be more
difficult than that of most other $Z'$ states within \es~\cite{LR,LRR}.  One
can see this feature in the relatively weak limits placed on $Z_I$ production
in present Tevatron data \cite{CDFZp}.  A $Z_I$ of 511 GeV (corresponding to
the highest-mass $e^+ e^-$ pair observed by CDF) is a possible candidate for
such a state.

The branching ratios for $Z_I$ decay can be deduced from the states with
masses below $M(Z_I)/2$ with $I_{3I} = \pm 1/2$, as in Eq.~(\ref{eqn:decomp}).
Thus, for three such families, the branching ratio to $e^+ e^-$ would be
$1/36 \simeq 3\%$, not very different from that of a standard $Z$.  The
presence of superpartners in final states would lower branching ratios
further \cite{sp}.

The subprocess $d \bar d \to Z_I \to e^- e^+$ is characterized by a maximal
angular asymmetry (i.e., $A_{FB} = -3/4$) in the backward direction \cite{FBA},
as one can see from the couplings in Eq.~(\ref{eqn:decomp}). This is in
contrast to the large forward asymmetry $A_{FB} \simeq 0.6$ expected \cite{FBA}
and observed \cite{CDFas} for the subprocesses $(u \bar u~{\rm or} ~d \bar d)
\to (\gamma^*,Z^*) \to e^- e^+$ in the standard model for $e^- e^+$ masses in
the Drell-Yan continuum well above the $Z$. 

The $Z_I$ can decay to pairs of exotic fermions such as $h \bar h$, $\nu_E \bar
\nu_E$, $E^- E^+$, $\bar N_e N_e$, and $n_e \bar n_e$.  It thus acts as a
gateway from the conventional world to exotic matter, allowing the production
of higher-mass states (or states produced with more transverse momentum) than
the conventional Drell-Yan processes involving virtual photons, $Z$'s, or
gluons.
\newpage

\leftline{\bf B.  Processes mediated by $W_I$ exchange}
\bigskip

Every member of one column of the $({\bf 2}_I,{\bf 6}^*)_L$ multiplet in
Eq.~(\ref{eqn:decomp}) can couple to the corresponding member of the other
column through emission or absorption of a (probably virtual) $W_I$. In some
cases, as in top quark decay, the gauge boson which mediates the decay may even
be on its mass shell. There thus arises the possibility of a new class of beta
decays, whose details depend on the combined masses of various doublets of
SU(2)$_I$. 

We have argued that the states $n$ are likely to be fairly light.  One
possibility for the end-product of decays mediated by $W_I$ exchange is for
them to involve $\bar N \bar n$ pairs.  This mechanism will make sense if $\bar
N$ does not acquire too large a Majorana mass, or is somehow prevented from
acquiring a Dirac mass in combination with $\nu$.  A means must then be found
for the $\bar N$ to decay.  This may take place through a radiative mechanism,
such as $\bar N \to \gamma n$.  Such processes can arise as a result of loop
diagrams involving mixing \cite{Rizzo,EM}.  The lifetime must be sufficiently
short that the decay occurs within the detector (so that photons are
detected), but not short enough to imply large flavor-changing neutral
currents, on which there are stringent constraints \cite{FCNC}.

An alterative ``lightest pair'' would be $\nu_E \bar \nu_e$.  In that case it
would be the $\nu_E$ which would have to undergo radiative decay, perhaps
to $\gamma \nu_e$.

Box diagrams involving $W_I$ exchange and intermediate $h$-type quarks can lead
to effective flavor-changing neutral interactions of the right-handed $d,~s$,
and $b$ quarks or their left-handed antiquark counterparts (as these are the
ones in \sui~doublets).  The suppression of these interactions below the levels
of ordinary flavor-changing neutral interactions induced by SU(2)$_L$
interactions imposes constraints on the CKM-like matrix describing the
\sui~couplings between $d,~s,~b$ and the corresponding $h$-type quarks.  These
appear to be easily satisfied for $h$-type quarks no heavier than the top quark
and $W_I$ masses in the range of several hundred GeV.  A more serious
constraint could in principle arise from the process $\mu \to e \gamma$, which
can be mediated by loops involving a $W_I$ and an intermediate exotic charged
lepton. Retracing steps taken in \cite{JandR}, it turns out that with
reasonable assumptions about mixing between light and heavy leptons this
process is predicted to occur at a rate below present limits. 
\bigskip

\leftline{\bf C.  Interpretation of the CDF \evt~event}
\bigskip

One event of the form $p \bar p \to$ \evt $ + \ldots$ (event 257646 of run
68739) has been reported at $\sqrt{s} = 1.8$ TeV by the CDF \cn~at the Fermilab
Tevatron \cite{eegg}.  A possible interpretation of this event is the
production of an $E^- E^+$ pair via the subprocess $d \bar d \to Z_I \to E^-
E^+$ (which has a maximal negative forward-backward asymmetry $A_{FB} = -3/4$,
just like $d \bar d \to Z_I \to e^- e^+$).  The $E^\pm$ states then decay to
$e^\pm$ and virtual (or perhaps real) $W_I$'s, which then materialize into
whatever doublets of SU(2)$_I$ are energetically accessible (such as the
possibilities mentioned above).  The decays of virtual $W_I$'s are thus
conceivable sources of photons + (missing energy) in a wide class of events. 

A likely mass for $E$ lies between the maximum beam energy currently attained
by LEP (80.5 GeV) and slightly below half the mass of the $Z_I$ candidate
mentioned above (511 GeV/2 $\simeq$ 250 GeV).  Depending on the masses of the
other exotic fermions, the $Z_I$ could decay to a number of pairs of such
states, including exotic charged leptons which we may call $M$ and $T$ of the
second and third families, $h \bar h$ (for one or more families) and the
\sui-doublet exotic neutral leptons [see Eq.~(\ref{eqn:decomp})].  At the very
least, one should expect to see at least one $\nu_E \bar \nu_E$ pair, most
likely leading to a pair of photons and missing energy as discussed below in
Sec.~IV A. 
\bigskip

\leftline{\bf D.  Scalar particles}
\bigskip

The existence of an extended Higgs structure within \es, based on bosons
belonging to the {\bf 27}-plet, implies that in addition to the neutral bosons
noted in Table I there are likely to be some light scalars with electromagnetic
charges $Q = \pm 1$.  (Some of the corresponding colored scalars can mediate
proton decay and must be very heavy \cite{DG}.)  We mention this possibility
only to note how rich the \es~spectrum is likely to be; to demonstrate that it
is evidence for supersymmetry may require considerable effort, such as the
comparison of couplings with one another. 
\bigskip

\leftline{\bf E.  Other signatures in hadron collisions}
\bigskip

The exchange of virtual $W_I$ quanta can lead to the production of pairs of
exotic quarks through the process $d \bar d \to h \bar h$ at subenergies
below that where direct $Z_I$ production can contribute \cite{LR}.  Whether
through $W_I$ exchange or via $Z_I$ in the direct channel, the angular
asymmetry of the subprocess should be maximal (i.e., $A_{FB} = 3/4$) in the
forward direction.  The decays of $h$ and $\bar h$ will be similar to those of
$E^+$ and $E^-$, but with down-type quarks replacing charged leptons. 

Production of $\nu_E \bar \nu_E$ pairs through $Z_I$ decay should lead to
pairs of photons + (missing energy) if the major decay modes of $\nu_E$
are radiative or involve a radiative chain.

It may be that decays like $E^- \to \nu_E + (\ldots)^-$ can compete favorably
with decays mediated by $W_I$.  In that case the system $(\ldots)$ can be any
decay product of a (probably virtual) $W^-$, and may include hadron jets as
well as leptons of any flavor.  However, if a large weak-isosinglet Dirac mass
is induced for both $E$ and $\nu_E$, these two states may be fairly close to
one another in mass. 
\bigskip

\leftline{\bf F.  CDF trilepton event}
\bigskip

Another exotic event (run 67581 / event 129896) reported by the CDF
Collaboration \cite{eegg} involves an $e^+ e^-$ pair, a $\mu^-$, a jet, and
missing transverse energy.  This could be due to $Z_I \to E^+ E^-$, where the
decays of $E^\pm$ lead to subsequent $e^\pm$ pairs, possibly through chains of
ordinary weak charge-changing transitions.  The muon and missing energy might
be the decay products of one such (perhaps virtual) $W$, while the jet might be
the (merged) decay products of another. 
\bigskip

\centerline{\bf IV.  OTHER COLLIDERS}
\bigskip

\leftline{\bf A.  Electron-positron colliders}
\bigskip

The reaction $e^+ e^- \to Z_I \to \ldots$ is an obvious gateway to new physics.
However \cite{LR}, one can also expect an observable rate for $W_I$ exchange
in the process $e^+ e^- \to E^+ E^-$ even at energies not corresponding to
$Z_I$ formation in the direct channel.  Moreover, all the exotic fermions with
the exception of $\bar N_e$ and $n_e$ can be produced via virtual photons
and/or $Z$'s in the direct channel.

Define $x \equiv \sin^2 \theta$, $s \equiv E_{\rm c.m.}^2$, and $r \equiv
[s/(s-m_Z^2)x(1-x)]$.  Then far from the $Z$ pole, where the $Z$ width can be
neglected, the contribution of a virtual photon and $Z$ in the direct channel
to the cross section for production of a fermion with electric charge $Q_f$ and
axial and vector $Z$ couplings $g_A$ and $g_V$ is 
\beq
\sigma(e^+ e^- \to f \bar f) = \sigma_\gamma \left\{
Q_f^2 - 2 r Q_f g_V^e g_V^f + r^2 [ (g_V^e)^2 + (g_A^e)^2] [(g_V^f)^2 +
\frac{\beta^2}{K_V}(g_A^f)^2] \right\}~~,
\eeq
where
\beq
\sigma_\gamma \equiv \frac{4 \pi \alpha^2}{3 s} N_c \beta K_V~,~~~
\beta \equiv \left( 1 - \frac{4 m_f^2}{s} \right)^{1/2}~,~~~
K_V \equiv \frac{3 - \beta^2}{2}~~,
\eeq
and $N_c$ is the number of colors of fermions.  For quarks ($N_c = 3$) the
cross section should be multiplied by an additional correction factor of $1 +
(\alpha_s/\pi) \approx 1.04$.  The values of $\sso$ far above pair production
threshold, where $\sigma_0 \equiv \sigma(e^+ e^- \to \gamma^* \to \mu^+
\mu^-$), are compared in Table II for various fermion species $f$ when the
energy is far below the $Z$ pole (where only the virtual photon dominates) and
when it is far above the $Z$ (where the interference in vector contributions of
the photon and $Z$ is possible). In computing the values of $g_V$ and $g_A$ for
$E^-$ and a Dirac neutrino $\nu_E$ one must recall that both left-handed and
right-handed states have the same value of $I_{3L}$: $-1/2$ for $E^-$ and
$+1/2$ for $\nu_E$. 

\begin{table}
\caption{Cross sections $\sigma$ [in units of $\sigma_0 \equiv \sigma(e^+ e^-
\to \gamma^* \to \mu^+ \mu^-$)] for $e^+ e^-$ production of pairs of fermions
$f \bar f$ via virtual photons and $Z$'s in the direct channel.  Here
$t$-channel exchanges are neglected for $e$ and $\nu_e$. The $\nu_E$ is assumed
to be a Dirac neutrino. Values of $g_V^f$ are quoted for $x = 0.2315$.  QCD
corrections to quark production have been neglected.} 
\begin{center}
\begin{tabular}{c c c c c c} \hline
Fermion & $Q_f$  & $g_V^f$   & $g_A^f$ &$\sso$ far &$\sso$ far \\
  $f$   &        &           &         & below $Z$ & above $Z$ \\ \hline
  $u$   & ~~2/3  & ~~0.0957  &  $-1/4$ &   4/3     &   1.80    \\
  $d$   & $-1/3$ & $-0.1728$ &   1/4   &   1/3     &   0.92    \\
  $e^-$ &  $-1$  & $-0.0185$ &   1/4   &    1      &   1.13    \\
$\nu_e$ &    0   &    1/4    &  $-1/4$ &    0      &   0.25    \\
  $h$   & $-1/3$ & ~~0.0772  &    0    &   1/3     &   0.35    \\
  $E^-$ &  $-1$  & $-0.2685$ &    0    &    1      &   1.20    \\
$\nu_E$ &    0   &    1/2    &    0    &    0      &   0.50    \\ \hline
\end{tabular}
\end{center}
\end{table}

All the exotic fermions $h$, $E$, and $\nu_E$ (assuming the last is a Dirac
particle) are produced exclusively via their vector couplings, and so are
excited with a cross section which attains its maximum not far above the
threshold energy $E_{\rm th}$.  The peak occurs at the maximum value of
$\beta(3-\beta^2)(1-\beta^2)$, or $E_{\rm c.m.} = 1.18 E_{\rm th}$ for very
heavy fermions, but somewhat lower when the ratio $M_Z/2m_f$ is non-negligible
as a result of the proximity of the $Z$ pole.  Thus, for example, for Dirac
neutrinos with $m(\nu_E) = 70$, 80, 90 GeV the respective cross sections for
$e^+ e^-\to \nu_E \bar \nu_E$ peak at 2.9, 1.8, and 1.2 pb for $E_{\rm c.m.} =
154$, 179, and 204 GeV, which are 1.10, 1.12, and 1.13 times $E_{\rm th}$.

With our present interpretation of the CDF \evt~event, the lowest-energy
signature for new physics in an electron-positron collider (such as LEP) could
be the process $e^+ e^- \to Z^* \to \nu_E \bar \nu_E$, followed by the
radiative decay of each $\nu_E$ to $\gamma n_e$.  In this case, one would see
events with two non-coplanar photons whose energies would become more and more
monochromatic as the machine energy was lowered toward $\nu_E \bar \nu_E$
threshold.  Such a signature is also a feature of neutralino pair production in
several supersymmetric scenarios \cite{sseegg}.  On the other hand, if it is
the $\bar N_e$ and not the $\nu_E$ which is undergoing radiative decay, the
reaction $e^+ e^- \to \nu_E \bar \nu_E$ may still act as a gateway to the
production of pairs of acoplanar photons, but their energies will not be
monochromatic even at $\nu_E \bar \nu_E$ threshold since they will then be
produced via the chain 
\beq
\nu_E \to \nu_e W_I^* \to \nu_e \bar N_e \bar n_e
\to \nu_e n_e \gamma \bar n_e~~~.
\eeq

\bigskip

\leftline{\bf B.  Electron-proton collisions}
\bigskip

In electron-proton collisions, the subprocess $e^- d \to E^- h$ is allowed by
$W_I$ exchange \cite{LR}.  The subprocess $e^+ d \to E^+ h$ involves a mismatch
of \sui~quantum numbers and is forbidden.  Thus, at the HERA collider, $e^- p$
collisions afford a better chance than $e^+ p$ collisions for discovering the
new fermions proposed here.  As in other experiments, one signature for new
physics would be the observation of events with isolated photons and missing
transverse energy.
\bigskip

\centerline{\bf V.  CONCLUSIONS}
\bigskip

We have investigated some features of the symmetry chain \es~$\to$ \sub~which
illustrate the richness of the group \es~ for exhibiting new physics at
present-day colliders.  An ``inert'' SU(2) subgroup, involving one $Z_I$ and
two $W_I$ bosons, can manifest itself through direct production of the $Z_I$,
production of exotic fermions, and decays of these fermions which can proceed
through several chains before ending up in a radiative cascade.  The present
scenario is thus one which lends itself to interpretation of an event involving
an \evt~final state reported by the CDF \cn~at Fermilab.  The favored
interpretation is 
\beq
\bar p p \to Z_I + \ldots \to E^+ E^- + \ldots
\eeq
followed by the chain
\beq
E^- \to e^- W_I^{(*)} \to e^- \bar N_e \bar n_e \to e^- \gamma n_e \bar n_e
\eeq
and its charge-conjguate for $E^+$ decay.  The $n_e$ state is allowed to be
stable as long as its mass satisfies cosmological bounds (typically less than a
few tens of eV).  The $Z_I$ is a neutral gauge boson with mass greater than
present limits \cite{CDFZp} of a few hundred GeV.  The $W_I$ is probably
virtual, as indicated by the asterisk in parentheses. The neutral nature of all
three bosons in \sui~is a key feature permitting the flavor of $E^-$ to be
passed on to the electron. 

Implications of the present \es~scheme include: (1) the expectation of $\gamma
\gamma$ events with missing energy but no charged lepton pairs, both in
proton-antiproton collisions at $E_{\rm c.m.} = 1.8$ TeV and in
electron-positron annihilations at sufficiently high energy, (2) the
confirmation of other decay modes of the ``gateway'' state $Z_I$, and (3)
the possibility of $W_I$-exchange processes in a number of reactions such
as electron-proton collisions, leading to pair-production of exotic states.

The purpose of this exercise was in part to see if the CDF event could be viewed
in a manner other than that involving supersymmetry \cite{sseegg} (see also
\cite{BM}).  This being said, the present story has several features in
common with the supersymmetric versions.  One may, in fact, have to work
rather hard to demonstrate whether the phenomena described above are really an
alternative to supersymmetry, or evidence for it. 

\begin{itemize}

\item The grand unified group is SU(5).  One cannot invoke multi-scale symmetry
breaking to obtain satisfactory predictions for the weak mixing angle or proton
decay.  The matter spectrum associated with supersymmetry provides a
satisfactory description within SU(5), but it remains to be seen whether the
spectrum of fermions and Higgs representations proposed here (which may be only
part of a supersymmetric spectrum) can do as well. 

\item The exotic leptons look somewhat like charginos (or selectrons) and
neutralinos, which also can decay via chains involving missing energy and
photons.  The missing transverse energy in the event (around 53 GeV) when
compared to the average transverse energy of the observed photons and leptons
(around 41 GeV), is more characteristic of a pair of missing particles as in
the supersymmetry scenario than of the two $n_e \bar n_e$ pairs implied by the
present scheme. (We are using a statistical estimate whereby 53/41 is closer to
$\sqrt{2}$ than to $\sqrt{4}$.) 

\item The use of {\bf 27}-plet multiplets of \es~both for matter (fermions) and
Higgs particles (bosons) is an invitation to make the theory supersymmetric. 
On the other hand, we have not made the gauge sector supersymmetric; we have
not necessarily invoked selection rules like R-parity which distinguish
superpartners from ordinary particles; and we have not required the existence
of three {\bf 27}-plets of Higgs bosons as superpartners for our three {\bf
27}-plets of fermions. 

\end{itemize}

The pattern of quarks and leptons has been quite regular up to now, just as if
the periodic table of the elements consisted only of rows of equal length and
were missing hydrogen, helium, the transition metals, the lanthanides, and the
actinides.  The new heavy states proposed here are the particle analogues of
the transition metals. The light ones could be the analogues of hydrogen and
helium.  Such new states could help us to make sense of the pattern of the
masses of the more familiar ones. 

\section*{Acknowledgments} 

I thank the CERN Theory Group for hospitality during this study, and G.
Alexander, P. Frampton, H. Frisch, D. London, M. Mangano, M. Schmitt, S. C. C.
Ting, and M. Veltman for fruitful discussions. This work was supported in part
by the United States Department of Energy under Contract No.~DE FG02 90ER40560.
 
\def \ajp#1#2#3{Am.~J.~Phys.~{\bf#1}, #2 (#3)}
\def \apny#1#2#3{Ann.~Phys.~(N.Y.) {\bf#1}, #2 (#3)}
\def \app#1#2#3{Acta Phys.~Polonica {\bf#1}, #2 (#3)}
\def \arnps#1#2#3{Ann.~Rev.~Nucl.~Part.~Sci.~{\bf#1}, #2 (#3)}
\def \baps#1#2#3{Bull.~Am.~Phys.~Soc. {\bf#1}, #2 (#3)}
\def \cmp#1#2#3{Commun.~Math.~Phys.~{\bf#1}, #2 (#3)}
\def \cmts#1#2#3{Comments on Nucl.~Part.~Phys.~{\bf#1}, #2 (#3)}
\def \cn{Collaboration}
\def \corn93{{\it Lepton and Photon Interactions:  XVI International Symposium,
Ithaca, NY August 1993}, AIP Conference Proceedings No.~302, ed.~by P. Drell
and D. Rubin (AIP, New York, 1994)}
\def \cp89{{\it CP Violation,} edited by C. Jarlskog (World Scientific,
Singapore, 1989)}
\def \dpff{{\it The Fermilab Meeting -- DPF 92} (7th Meeting of the American
Physical Society Division of Particles and Fields), 10--14 November 1992,
ed. by C. H. Albright \ite~(World Scientific, Singapore, 1993)}
\def \dpf94{DPF 94 Meeting, Albuquerque, NM, Aug.~2--6, 1994}
\def \efi{Enrico Fermi Institute Report No. EFI}
\def \el#1#2#3{Europhys.~Lett.~{\bf#1}, #2 (#3)}
\def \f79{{\it Proceedings of the 1979 International Symposium on Lepton and
Photon Interactions at High Energies,} Fermilab, August 23-29, 1979, ed.~by
T. B. W. Kirk and H. D. I. Abarbanel (Fermi National Accelerator Laboratory,
Batavia, IL, 1979}
\def \hb87{{\it Proceeding of the 1987 International Symposium on Lepton and
Photon Interactions at High Energies,} Hamburg, 1987, ed.~by W. Bartel
and R. R\"uckl (Nucl. Phys. B, Proc. Suppl., vol. 3) (North-Holland,
Amsterdam, 1988)}
\def \ib{{\it ibid.}~}
\def \ibj#1#2#3{~{\bf#1}, #2 (#3)}
\def \ichep72{{\it Proceedings of the XVI International Conference on High
Energy Physics}, Chicago and Batavia, Illinois, Sept. 6--13, 1972,
edited by J. D. Jackson, A. Roberts, and R. Donaldson (Fermilab, Batavia,
IL, 1972)}
\def \ijmpa#1#2#3{Int.~J.~Mod.~Phys.~A {\bf#1}, #2 (#3)}
\def \ite{{\it et al.}}
\def \jmp#1#2#3{J.~Math.~Phys.~{\bf#1}, #2 (#3)}
\def \jpg#1#2#3{J.~Phys.~G {\bf#1}, #2 (#3)}
\def \lkl87{{\it Selected Topics in Electroweak Interactions} (Proceedings of
the Second Lake Louise Institute on New Frontiers in Particle Physics, 15--21
February, 1987), edited by J. M. Cameron \ite~(World Scientific, Singapore,
1987)}
\def \ky85{{\it Proceedings of the International Symposium on Lepton and
Photon Interactions at High Energy,} Kyoto, Aug.~19-24, 1985, edited by M.
Konuma and K. Takahashi (Kyoto Univ., Kyoto, 1985)}
\def \mpla#1#2#3{Mod.~Phys.~Lett.~A {\bf#1}, #2 (#3)}
\def \nc#1#2#3{Nuovo Cim.~{\bf#1}, #2 (#3)}
\def \np#1#2#3{Nucl.~Phys.~{\bf#1}, #2 (#3)}
\def \pisma#1#2#3#4{Pis'ma Zh.~Eksp.~Teor.~Fiz.~{\bf#1}, #2 (#3) [JETP Lett.
{\bf#1}, #4 (#3)]}
\def \pl#1#2#3{Phys.~Lett.~{\bf#1}, #2 (#3)}
\def \plb#1#2#3{Phys.~Lett.~B {\bf#1}, #2 (#3)}
\def \pra#1#2#3{Phys.~Rev.~A {\bf#1}, #2 (#3)}
\def \prd#1#2#3{Phys.~Rev.~D {\bf#1}, #2 (#3)}
\def \prl#1#2#3{Phys.~Rev.~Lett.~{\bf#1}, #2 (#3)}
\def \prp#1#2#3{Phys.~Rep.~{\bf#1}, #2 (#3)}
\def \ptp#1#2#3{Prog.~Theor.~Phys.~{\bf#1}, #2 (#3)}
\def \rmp#1#2#3{Rev.~Mod.~Phys.~{\bf#1}, #2 (#3)}
\def \rp#1{~~~~~\ldots\ldots{\rm rp~}{#1}~~~~~}
\def \si90{25th International Conference on High Energy Physics, Singapore,
Aug. 2-8, 1990}
\def \slc87{{\it Proceedings of the Salt Lake City Meeting} (Division of
Particles and Fields, American Physical Society, Salt Lake City, Utah, 1987),
ed.~by C. DeTar and J. S. Ball (World Scientific, Singapore, 1987)}
\def \slac89{{\it Proceedings of the XIVth International Symposium on
Lepton and Photon Interactions,} Stanford, California, 1989, edited by M.
Riordan (World Scientific, Singapore, 1990)}
\def \smass82{{\it Proceedings of the 1982 DPF Summer Study on Elementary
Particle Physics and Future Facilities}, Snowmass, Colorado, edited by R.
Donaldson, R. Gustafson, and F. Paige (World Scientific, Singapore, 1982)}
\def \smass90{{\it Research Directions for the Decade} (Proceedings of the
1990 Summer Study on High Energy Physics, June 25 -- July 13, Snowmass,
Colorado), edited by E. L. Berger (World Scientific, Singapore, 1992)}
\def \stone{{\it B Decays}, edited by S. Stone (World Scientific, Singapore,
1994)}
\def \tasi90{{\it Testing the Standard Model} (Proceedings of the 1990
Theoretical Advanced Study Institute in Elementary Particle Physics, Boulder,
Colorado, 3--27 June, 1990), edited by M. Cveti\v{c} and P. Langacker
(World Scientific, Singapore, 1991)}
\def \yaf#1#2#3#4{Yad.~Fiz.~{\bf#1}, #2 (#3) [Sov.~J.~Nucl.~Phys.~{\bf #1}, #4
(#3)]}
\def \zhetf#1#2#3#4#5#6{Zh.~Eksp.~Teor.~Fiz.~{\bf #1}, #2 (#3) [Sov.~Phys. -
JETP {\bf #4}, #5 (#6)]}
\def \zpc#1#2#3{Z.~Phys.~C {\bf#1}, #2  (#3)}

\end{document}